\documentstyle[aps,prb,epsfig,amssymb,floats]{revtex}

\begin{document}
\draft

\twocolumn[\hsize\textwidth\columnwidth\hsize\csname
@twocolumnfalse\endcsname

\title{Amorphous ZrO$_2$ from {\em Ab-initio} molecular dynamics:\\
Structural, electronic and dielectric properties}

\author{Xinyuan Zhao \cite{pad}, Davide Ceresoli, and David Vanderbilt}
\address{Department of Physics and Astronomy, Rutgers University,
	Piscataway, NJ 08854-8019}
\date{October 13, 2004}
\maketitle
\begin{abstract}
Realistic models of amorphous ZrO$_2$ are generated in a
``melt-and-quench'' fashion using {\em ab-initio} molecular
dynamics in a plane-wave pseudopotential formulation of
density-functional theory.  The structural properties of the
resulting amorphous models are analyzed, with special attention to
coordination statistics. The vibrational and dielectric properties
of one of these models are then investigated from first principles
using linear-response methods.  The electronic dielectric constant
and Born effective charges are found to be very similar to those
of the crystalline phases.  Encouragingly, the predicted total static
dielectric constant is $\sim$22, comparable to that of the monoclinic
phase.  This work is motivated by the
search for improved gate dielectric materials for sub-0.1\,$\mu$m CMOS
technology, and may also have implications for HfO$_2$ and for
silicates of ZrO$_2$ and HfO$_2$.
\end{abstract}


\vskip2pc]

\columnseprule 0pt
\narrowtext

\section{Introduction}
\label{sec:intro}

The projections in the International Technology Roadmap for
Semiconductors call for an effective gate dielectric thickness of
approximately 1\,nm for 0.1\,$\mu$m CMOS technology by the year
2006. However, the use of current SiO$_2$ and silica nitride
materials in this regime will become problematic because of
intolerably high leakage currents. A possible solution, currently
under intensive exploration, is to replace SiO$_2$ as the CMOS gate
dielectric by an oxide having a much higher dielectric constant.
Use of such a ``high-$K$ dielectric'' would allow one to make
physically thicker films, hence reducing the leakage current, while
at the same time maintaining or even increasing the gate
capacitance.  High-$K$ metal oxides are at the focus of this
effort, with HfO$_2$, ZrO$_2$, and their mixtures with SiO$_2$ showing
great promise.\cite{wilk,gusev}

In previous work, we studied the structural, electronic, and
lattice dielectric propterties of crystalline phases of ZrO$_2$ and
HfO$_2$ (Refs.~\onlinecite{zro2,hfo2,ortho,zrhf}) using a
first-principles density-functional approach.  We found that the
lattice dielectric constant depends strongly on crystal phase; for
example, for ZrO$_2$, we find orientationally-averaged total
$\epsilon$ values of 37 and 38 for cubic and tetragonal phases, and
smaller values of 20, 20, and 19 for the monoclinic and for two
orthorhombic phases, respectively.  We also found strong
anisotropies, most notably a large in-plane susceptibility for the
tetragonal phase.  Differences in the electronic density of states
and in band gaps were also studied,\cite{ortho,zrhf} but these are
not profound enough to have much effect on the purely electronic
dielectric constant, which remains approximately 5 for all crystal
phases.  These results suggest that novel structural modifications
of ZrO$_2$ or HfO$_2$ might possibly provide a route to enhanced
dielectric constants.

One of the great advantages of using SiO$_2$ for CMOS technology
has been the fact that it forms an {\it amorphous} oxide, thus
allowing it to conform to the substrate with enough freedom to
eliminate most electrical defects at the interface.  Materials
like ZrO$_2$ and HfO$_2$ tend to crystallize much more readily than
SiO$_2$, raising questions about whether equally smooth and clean
interfaces can be formed from such materials.  While it is possible to
form such oxides in an amorphous state by low-temperature processes,
they tend to recrystallize during the thermal treatments needed for
later stages of device processing.  One current
avenue of investigation is to consider crystalline oxides, such as
perovskites, that can be grown epitaxially on Si.\cite{mckee}
Another is to search for ways to raise the recrystallization
temperature of ZrO$_2$- or HfO$_2$-based materials, e.g., by
incorporating Si, Al, or N into the random network structure.
However, very little theoretical guidance is available for such
initiatives, since almost nothing is known about the structure and
atomic-scale properties even for pure amorphous ZrO$_2$ or HfO$_2$.

With these motivations, we have embarked on a theoretical study of
the structure and properties of amorphous ZrO$_2$ ($a$-ZrO$_2$).  In
this paper, we report the construction of realistic models of
$a$-ZrO$_2$ through ``melt-and-quench'' {\em ab-initio}
molecular-dynamics (MD) simulations.  This approach, first
pioneered by Car and Parrinello,\cite{cpmd} combines
density-functional theory (DFT) with MD into a powerful tool for
investigating the physics of large systems, especially for liquids
and amorphous structures where the atomic coordinates cannot be
obtained from diffraction experiments.  Using the
generated models as prototypes for $a$-ZrO$_2$, we investigate the
atomistic structure, especially the bonding and coordination statistics,
and then focus in detail on the
lattice dielectric properties.  While the calculations have been
carried out only for $a$-ZrO$_2$, it can be expected that many
conclusions would hold also for $a$-HfO$_2$ in view of the close
chemical and crystallographic similarity of these two materials.

The paper is organized as follows.  Section \ref{sec:thy} describes the
theoretical approach of
our first-principles simulations, the supercells used for modeling $a$-ZrO$_2$,
and the linear-response method used to analyze the lattice dielectric
properties. In Sec.~\ref{sec:md} we present
and discuss the structural properties of the amorphous models obtained
from the MD simulations, with special attention to bonding and coordination.
The results of the linear-response calculations, such as the Born effective
charges, phonon modes, and dielectric tensors, are then presented and
discussed in Sec.~\ref{sec:diel}. Section \ref{sec:summary} concludes the paper.

\section{Theory}
\label{sec:thy}

\subsection{Details of static structural calculations}
\label{sec:thy_struct}

In this work, we perform {\em ab-initio} constant-temperature MD
as implemented in the VASP simulation package, \cite{vasp} in which
the electronic structure is described within DFT \cite{hk64,ks65}
in the local-density approximation (LDA), \cite{lda} using a
plane-wave basis and ultrasoft pseudopotentials. \cite{uspp} At
each molecular-dynamics step, the instantaneous Kohn-Sham electronic
wavefunctions are obtained using the RMM-DIIS (residual
minimization with directive inversion in the iterative subspace)
method,\cite{pulay80,wood85} which is found to be
particularly efficient for diagonalizing the Kohn-Sham Hamiltonian for
large systems such as the 96-atom supercell considered here.   The
forces at each MD step are calculated as the derivatives of
the generalized free energy with respect to the ionic
positions based on the Hellmann-Feynman theorem.\cite{hf-thm} The
calculated forces are then used to integrate Newton's equations of motion
for the ionic degrees of freedom via the Verlet algorithm.\cite{verlet}

\begin{table}
\caption{Structural parameters obtained for the three ZrO$_2$
phases from the present theory, compared with a previous theoretical
calculation of higher accuracy (see text).  Lattice parameters $a$, $b$,
and $c$ in ${\rm \AA}$; volume $V$ in ${\rm \AA}^{3}$; monoclinic angle
$\beta$ in degrees. }
\begin{center}
\begin{tabular}{ccdd}
Phase   & Parameter & This Work & Ref.~\protect\onlinecite{zro2}
\\ \hline
monoclinic & $a$     & 5.098   & 5.108 \\
           & $b$     & 5.171   & 5.169 \\
           & $c$     & 5.264   & 5.271 \\
           & $\beta$ & 99.49   & 99.21 \\
           & $V$     & 136.77  & 137.40 \\ \hline
tetragonal & $a$     & 5.037  & 5.029 \\
           & $c$     & 5.113  & 5.101 \\
           & $d_z$   & 0.041  & 0.042 \\
           & $V$     & 129.73 & 129.03 \\ \hline
cubic &      $a$     & 5.034  & 5.037 \\
      &      $V$     & 127.57 & 127.80 \\
\end{tabular}
\end{center}
\label{table:md_uspp}
\end{table}

\begin{figure}[t]
\begin{center}
   \epsfig{file=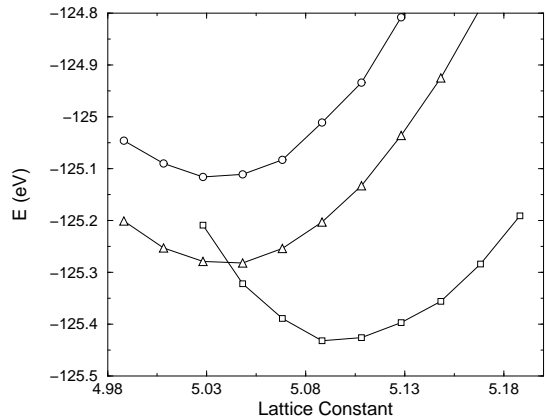,width=2.8in}
\end{center}
\caption{Relative energies per formula unit versus lattice constant
$a$ (see Table \protect\ref{table:md_uspp}) for the monoclinic ($\Box$),
tetragonal ($\bigtriangleup$), and cubic ($\circ$) phases of ZrO$_2$.}
\label{fig:md_ev}
\end{figure}

Because MD simulations on large supercells are quite computer-intensive,
we have taken some care to make the calculations as efficient as
possible.  In particular, we have reduced the plane-wave cutoff energy
to 15 Ry, and the pseudopotential for Zr includes only the outermost
shells (4$d$, 5$s$) in the valence.  To test the adequacy of these
approximations, we tabulate in
Table~\ref{table:md_uspp} the lattice parameters
calculated with VASP using the current settings and those determined
in Ref.~\onlinecite{zro2}, where the cutoff was 25 Ry and the
4$s$ and 4$p$ Zr levels were included in the valence.
It can be readily seen that they agree
very well. In Fig.~\ref{fig:md_ev}, we show the calculated ground-state
energies of the three ZrO$_2$ phases (monoclinic, tetragonal, and cubic).
The energy ordering (monoclinic lowest, cubic highest) is qualitatively
and semiquantitatively correct.  We thus conclude that the chosen
pseudopotentials and plane-wave cut-off are adequate to correctly
reproduce the energetics of the three ZrO$_2$ phases.

\subsection{Supercell structures}
\label{sec:thy_cell}

In our simulation, $a$-ZrO$_2$ is modeled to have periodic boundary
conditions with a cubic supercell containing 96 atoms.  This is the
same content as a 2$\times$2$\times$2 repetition of the monoclinic
unit cell, since each monoclinic cell contains 4 chemical formula
units (4 Zr atoms and 8 oxygen atoms).  In fact, the cubic lattice
vectors are only slightly distorted from the 2$\times$2$\times$2
monoclinic ones ($\beta=99.21^{\circ}$ and the lengths of three
lattice vectors differ only within $3\%$).\cite{zro2}  Thus,
we use as a starting point a structure in which the lattice vectors
are perfectly cubic but the fractional coordinates are those of the
relaxed monoclinic structure.  (This monoclinic structure remains
stable in this environment; that is, when it is relaxed under fixed
cubic lattice vectors, only small further displacements occur and
the symmetry of the atomic coordinates remains monoclinic.)  All
subsequent MD simulations are done using fixed cubic lattice
vectors for the supercell, and using single $k$-point sampling at the
$\Gamma$ point of the Brillouin zone.

\subsection{{\em Ab-initio} molecular dynamics}
\label{sec:thy_md}

The ``melt-and-quench'' simulation scheme proceeds by carrying out a
series of MD simulations, each at a fixed temperature
(i.e., in the canonical ensemble).
Temperature $T$ is sequentially increased, step by step, from
room temperature ($T_m$) to some hypothetical temperature ($T_{\rm max}$)
high enough to liquefy the system. At each step, the structure obtained 
from the previous MD simulation is used as the starting point for the
next one. 
The process is then reversed to quench the system back to $T_m$. 
For each MD simulation, the system is given enough
time to reach its thermal equilibrium in order to eliminate its correlation
to the previous structure. In the case of $a$-ZrO$_2$,
we let the system evolve at temperatures ranging from zero
to 4000\,K, well above the experimental melting point of 2980\,K.
The system is then rapidly quenched back to 300\,K, followed by a
relaxation to zero temperature of both atomic coordinates and lattice
vectors.

A canonical ensemble is realized using a Nos\'{e}
thermostat, which modifies the Newtonian MD
by introducing an additional degree of freedom such that
the total energy of the physical system is allowed to fluctuate.
\cite{nose,hoover} In the algorithm of Nos\'{e}, the extra degree of
freedom --- the Nos\'{e} mass $Q$ ---  controls the frequency of the
temperature fluctuations. In principle, the Nos\'{e} mass should be chosen
so that the induced temperature fluctuations show approximately the
same frequencies as the ``typical'' phonon frequencies of the considered
system. \cite{vguide} 
We know from Ref.~\onlinecite{zro2} that the phonon frequencies at the zone
center in monoclinic ZrO$_2$ are in the range of 100\,--\,748\,cm$^{-1}$,
corresponding approximately to 40\,--\,300\,fs in period.
Our experience shows that $Q$ is rather insensitive to the
designated temperature.  In the case of ZrO$_2$, $Q$ =
0.15\,--\,0.20 has been found to be suitable.
For $Q$ in this range, the characteristic period of the temperature
fluctuations is approximately 40\,fs, which  corresponds the high end of the
phonon spectrum for monoclinic ZrO$_2$.

It has to be emphasized that VASP currently supports only constant-volume
MD simulations, i.e., the volume of a unit cell is fixed during a
``melt-and-quench'' simulation. In order to study the influence of volume on
the resultant structure, we carry out a series of similar MD simulations
on supercells of different 
volumes, as will be discussed in details in Sec. \ref{sec:md_details}.

\subsection{Details of the linear-response calculations}
\label{sec:thy_rf}

The dielectric properties of the MD-simulated amorphous models of ZrO$_2$
are calculated by the specialized linear-response techniques as implemented
in ABINIT, \cite{abinit} a simulation package that is also based
on DFT with pseudopotentials and plane-wave expansion. ABINIT
has the capability to calculate response functions such as phonon
dynamical matrices, dielectric tensors, and Born effective charge
tensors via a variational approach to density-functional perturbation
theory \cite{dfpt1,dfpt2} in which phonon displacements and
static homogeneous electric fields are treated as perturbations.
Our linear-response calculations are performed within the
LDA \cite{lda} using the Perdew-Wang's parameterization. \cite{pw92}
The Brillouin zone is sampled only at the $\Gamma$ point.
Extended norm-conserving pseudopotentials \cite{teter} with
valence configurations of Zr(4$s$,4$p$,4$d$,5$s$) and O(2$s$,2$p$),
and a cutoff energy of 35.0 Hartree, are found to provide satisfactory
convergence.

\section{Amorphous Models of Z\lowercase{r}O$_2$}
\label{sec:md}

\subsection{Generation of amorphous structures}
\label{sec:md_details}

\begin{figure}
\begin{center}
   \epsfig{file=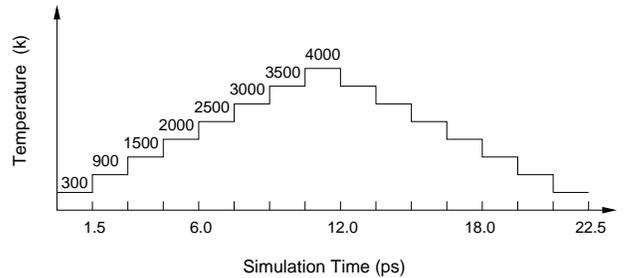,width=3.2in}
\end{center}
\caption{Programmed application of temperature vs.~time
in the ``melt-and-quench'' simulation.  Temperature is increased from 300 to
4000 K in steps of 1.5 ps, then decreased in a reversed sequence, for
a total simulation time of 22.5\,ps.}
\label{fig:mq}
\end{figure}

The ``melt-and-quench'' MD simulations are schematically illustrated 
in Fig. \ref{fig:mq}. At each selected temperature, the ZrO$_2$ system 
is allowed to run for 1.5\,ps to reach its thermodynamical equilibrium. 
The time step for the simulations is set to be 3 fs
for $T=$ 300\,--\,2500\,K, and is reduced to 2 fs 
when $T$ is in the range of 3000\,K to 4000\,K in order to
suppress numerical error at higher temperature. The total
simulation time is 22.5\,ps (see Fig. \ref{fig:mq}).

As mentioned in Sec.~\ref{sec:thy_md}, several supercells with
different volume are used in our simulation.  The prototypical
supercell, designated as the $p$-cell, is adjusted to have its volume
equal to $8\,V_{\rm mono}$, where $V_{\rm mono}$ is the volume of a unit
cell of monoclinic ZrO$_2$. The side length of
the $p$-cell is thus determined as 10.320\,a.u.. The other cells,
designated as the 2-, 4-,
7-, and $11$-cell, have their side lengths increased by 2\%, 4\%, 7\%,
and 11\% over that of the $p$-cell respectively. This information is
summarized in Table~\ref{table:md_cells}.

\begin{table}
\caption{Supercells used in the simulations. The side length
$a$ is in atomic units (a.u.). The percentage in the second column
indicates the increment of $a$ from the p-cell.}
\begin{tabular}{ccc}
Supercell Type & Increment (\%) & Side Length ($a$) \\
\tableline
$p$-cell & 0  & 10.320 \\
2-cell   & 2  & 10.527 \\
4-cell   & 4  & 10.727  \\
7-cell   & 7  & 11.043  \\
11-cell   & 11 & 11.467 \\
\end{tabular}
\label{table:md_cells}
\end{table}
\begin{table}[b]
\caption{Summary of structures resulting from the candidate
supercells.
$V_i$ is the volume at which the MD simulation is carried out;
$V_f$ is the volume after the final structural relaxation.
Fourth column shows the coordination number (CN),
or range of CNs, for Zr and O respectively.  Last column
characterizes the resulting structure.}
\begin{center}
\begin{tabular}{rcccc}
 & $V_i$ (\AA$^3$)   & $V_f$ (\AA$^3$)  & CN (Zr/O) & Descrip. \\ \hline
$p$-cell & 1099.2 & 1041.7 & 8 / 4     &  Crystal \\
2-cell   & 1166.5 & 1091.9 & \phantom{0--}7 / 3, 4 & Crystal \\
4-cell   & 1234.4 & 1040.8 & 8 / 4     & Crystal \\
7-cell   & 1346.1 & 1230.8 & 5--8 / 2--5 & Amorph. \\
11-cell  & 1507.9 & 1329.6 & -- & Collapsed  \\
\end{tabular}
\end{center}
\label{table:md_volume}
\end{table}

The ``melt-and-quench'' simulations are performed on these five candidate
supercells. The results are summarized in Table~\ref{table:md_volume}, 
where $V_i$ is the initial volume of the supercell that was kept fixed
during the MD simulation, and $V_f$ denotes the volume resulting from
the final zero-temperature structural relaxation of coordinates and
lattice vectors at the end of the simulation.
The coordination number of each atom is determined by counting the
number of atoms within a cutoff radius of 3.00\,\AA, which can be
compared with the range of Zr--O bond lengths, 2.035 -- 2.233\,\AA,
that characterizes monoclinic ZrO$_2$.\cite{zro2}
It can be readily seen that the volume has a significant effect on
the final structure. For the 11-cell, we find that the
atomistic structure collapses into smaller clusters accompanied with
``spatial voids'' even at room temperature ($T_m$), suggesting that
the supercell volume is too large to sustain the atomic structure.
For the $p$-, 2- and 4-cell, during the ``melt-and-quench'' simulation
the coordination numbers suggest some degree of disorder, but
the final structures recrystallize when the systems are quenched back and
relaxed to the ground state.

Fortunately, a reasonable amorphous model of ZrO$_2$ (called ``Model
I'') is realized from the 7-cell. Although it has an initial volume of
1346.1\,\AA$^3$, a subsequent relaxation to zero temperature
reduces the volume to 1230.8\,\AA$^3$, which is approximately
12\% bigger than $V_i$ of the $p$-cell. Using Model I as the starting
structure, we perform a second ``melt-and-quench'' simulation and
obtain another model of $a$-ZrO$_2$, which is called ``Model II''
hereafter. In the remainder of this section, we will mainly focus on
analyzing the structural properties of two models.

\subsection{Analysis of amorphous structures}
\label{sec:md_models}

\subsubsection{Amorphous Model I}
\label{sec:md_model1}

\begin{figure}
\begin{center}
   \epsfig{file=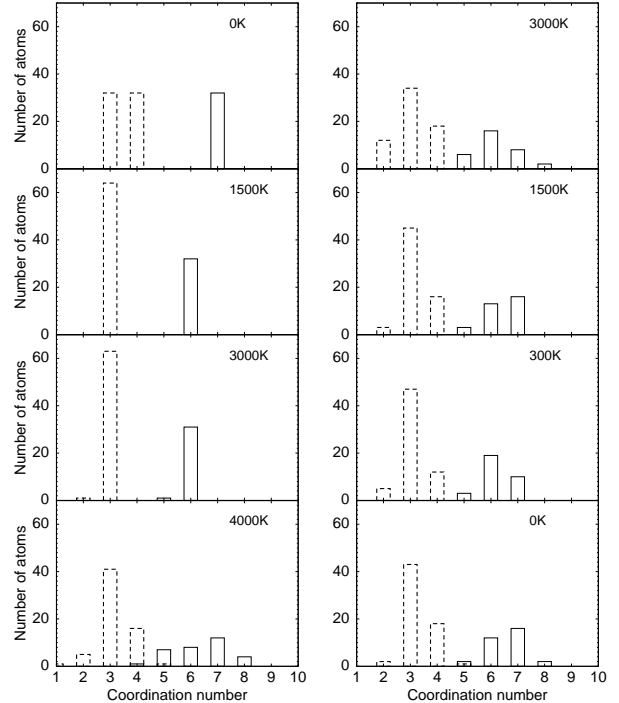,width=3.1in}
\end{center}
\caption{Distribution of coordination numbers during the
``melt-and-quench'' simulation giving rise to Model I. ``Melt'' and
``quench'' processes are shown at left and right, respectively.
Simulation temperature is indicated in each panel; Zr and O atoms
are indicated by solid and dashed bars, respectively.}
\label{fig:md_7mc_coor}
\end{figure}
\begin{figure}
\begin{center}
   \epsfig{file=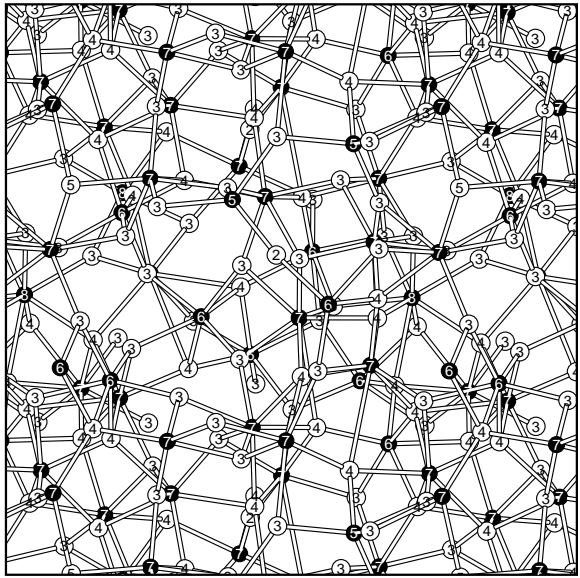,width=3.0in}
\end{center}
\caption{Structure of amorphous Model I.  Zr atoms are black; O atoms
are white.  Coordination number is indicated on each atom (for atoms
near front or back of view, some neighbors may not appear).}
\label{fig:md_q7mc}
\end{figure}

Figure \ref{fig:md_7mc_coor} tracks the atomic coordination number
(CN) of the Zr and O atoms during the formation of Model I, starting
from the 7-cell in which oxygen atoms have CN=3 or 4 and Zr
atoms have CN=7. Interestingly, the system retreats at $T$=1500\,K to a
higher-symmetry structure with CN=3 and 6 for O and Zr atoms
respectively. A hint of disorder is displayed at $T$=3000\,K via the
incidence of the 5-coordinated Zr atoms and the 2-coordinated O
atoms. When $T$ is increased to 4000\,K,
the distribution of coordination numbers suggests that the system
has become strongly disordered, which we take as a sign that it has
melted.  The system is then quenched quickly (about 12\,ps, see
Fig.~\ref{fig:mq}) back to $T_m$. From the variation
of the coordination numbers, one can readily see that the amorphous character
has survived to $T_m$. A snapshot of this  amorphous structure (Model I)
is shown in Fig.~\ref{fig:md_q7mc}, from which we can directly see
that a reasonable amorphous structure appears to have been generated.
The numbers 
on the atoms in Fig.~\ref{fig:md_q7mc} indicate the coordination numbers.

\begin{figure}
\begin{center}
   \epsfig{file=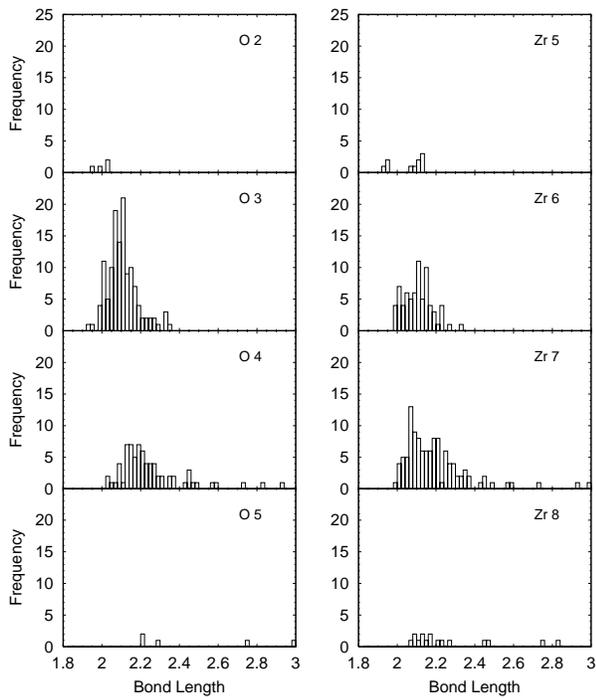,width=3.1in}
\end{center}
\caption{Bond length distributions in Model I.  Each panel shows
the number of bonds (frequency) {\em vs.}\ the bond length (\AA) for
a particular species (``O'' or ``Zr'') with a particular
coordination number (e.g., ``O$\,$4'' indicates 4-coordinated oxygen).}
\label{fig:md_bondstat_q7mc}
\end{figure}

The bond distributions of Zr and O atoms in Model I are plotted in
Fig.~\ref{fig:md_bondstat_q7mc}. While there are 5- and
8-coordinated Zr atoms in Model I, it is obvious that the 6- and
7-coordinated Zr atoms predominate. We can see from
Fig.~\ref{fig:md_bondstat_q7mc} that most of the oxygen atoms have
a CN of 3 or 4, except for a few 2- or 5-coordinated oxygen atoms.

\begin{table}
\caption{Bond (Zr-O) statistics in the amorphous models I and
II. Subscripts of Zr and O denote coordination numbers;
$N$ is the number of atoms having the specified coordination number.
The minimum ($L_{\rm min}$ ), maximum ($L_{\rm max}$), and
average ($\overline{L}$) bond length for each type of atom
is given in \AA.}
\begin{center}
\begin{tabular}{ccrccccrccc}
 & &\multicolumn{4}{c}{Model I} & &\multicolumn{4}{c}{Model II} \\
 & & $N$  & $L_{\rm min}$ & $L_{\rm max}$ & $\overline{L}$ & & $N$ &
 $L_{\rm min}$ & $L_{\rm max}$ & $\overline{L}$ \\ \hline
Zr$_5$ & & 2  & 1.90 & 2.12 & 2.04 & & - & - & - & - \\
Zr$_6$ & & 12 & 1.96 & 2.32 & 2.09 & & 10 & 1.92 & 2.66 & 2.11 \\
Zr$_7$ & & 16 & 1.97 & 2.96 & 2.17 & & 18 & 1.95 & 2.98 & 2.16 \\
Zr$_8$ & & 2  & 2.05 & 2.82 & 2.25 & & 4  & 2.02 & 2.94 & 2.24 \\ \hline
O$_2$  & & 2  & 1.94 & 2.02 & 1.98 & & 3  & 1.92 & 2.05 & 1.98 \\
O$_3$  & & 43 & 1.91 & 2.33 & 2.09 & & 32 & 1.92 & 2.95 & 2.10 \\
O$_4$  & & 18 & 2.01 & 2.92 & 2.23 & & 29 & 1.99 & 2.98 & 2.22 \\
O$_5$  & & 1  & 2.18 & 2.96 & 2.47 & & - & - & - & -\\
\end{tabular}
\end{center}
\label{table:md_bondlength}
\end{table}

The number of atoms having each CN, together with the minimum
($L_{\rm min}$), maximum ($L_{\rm max}$), and average
($\overline{L}$) bond length for each CN, are reported in
Table~\ref{table:md_bondlength} for both Model I and Model II.  The
bond lengths are found to be in the range of 2.04 -- 2.25\,\AA\ for
$\overline{L}$, generally similar to the range of 2.035 --
2.233\,\AA\ that we found previously for monoclinic
ZrO$_2$.\cite{zro2} We find that the average bond length
$\overline{L}$ decreases monotonically as the coordination number
increases, consistent with the usual expectation, based on chemical
principles, of an inverse correlation between CN and bond
strength.  The minima ($L_{\rm min}$) and maxima ($L_{\rm max}$) of
the bond lengths manifest similar behavior.

\subsubsection{Amorphous Model II}
\label{sec:md_model2}

\begin{figure}
\begin{center}
   \epsfig{file=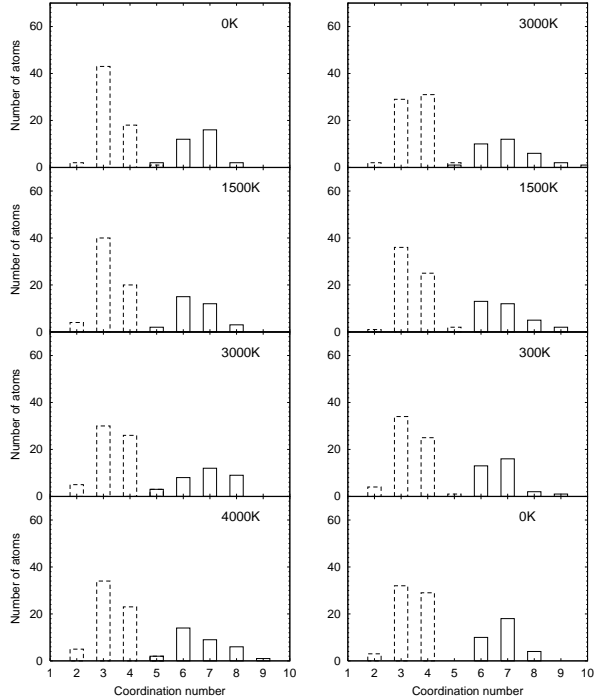,width=3.1in}
\end{center}
\caption{Distribution of coordination numbers during the
``melt-and-quench'' simulation giving rise to Model II. ``Melt'' and
``quench'' processes are shown at left and right, respectively.
Simulation temperature is indicated in each panel; Zr and O atoms
are indicated by solid and dashed bars, respectively.}
\label{fig:md_rel7_coor}
\end{figure}

\begin{figure}
\begin{center}
   \epsfig{file=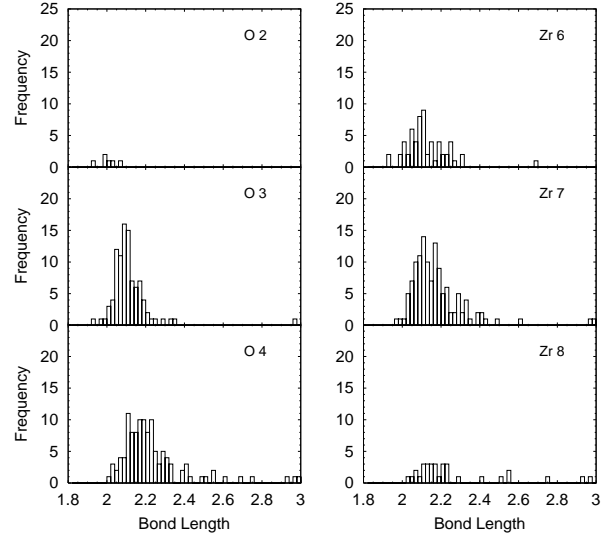,width=3.1in}
\end{center}
\caption{Bond length distributions in Model II.  Each panel shows
the number of bonds (frequency) {\em vs.}\ the bond length (\AA) for
a particular species (``O'' or ``Zr'') with a particular
coordination number.}
\label{fig:md_bondstat_rel7}
\end{figure}

As mentioned earlier, a second amorphous
structure, Model II, is obtained using Model I as the starting
input structure.  Figures \ref{fig:md_rel7_coor} and
\ref{fig:md_bondstat_rel7} present results for Model II in a
parallel fashion as for Figs.~\ref{fig:md_7mc_coor} and
\ref{fig:md_bondstat_q7mc} for Model I.  A significant change in
going from Model I to Model II is the lack of the 5-coordinated Zr
and O atoms (Zr$_5$ and O$_5$ in Table~\ref{table:md_bondlength}).
Interestingly, in Model II, the numbers of O$_3$ and O$_4$ atoms
have become nearly equal, while in Model I there was a
substantial difference ($N$=43 for O$_3$, 18 for O$_4$). The
relative populations of Zr$_6$ and Zr$_7$ do not change very much
from Model I to Model II, although a slight shift from Zr$_6$ to
Zr$_7$ has occurred. The average lengths of the bonds in the two
amorphous models are approximately equal. Finally, we point out
that the density of the system has increased by $\sim$6\% in going
from Model I to Model II.

\subsection{Discussion}
\label{sec:md_discussion}

We have presented, in this section, two models of amorphous ZrO$_2$
obtained via a ``melt-and-quench'' scheme using {\em ab-initio} MD.
Since the volume in our simulations is not allowed to vary in any single
MD run, certain candidate supercells of different volume have been constructed
in order to study the effects of volume on the final amorphous
structures (see Table~\ref{table:md_cells}). One candidate cell
(i.e. the 7-cell in Table~\ref{table:md_cells}) successfully leads
to an amorphous model of ZrO$_2$ (Model I). The density of the
7-cell is about 20\% less than the density of monoclinic
ZrO$_2$ ($\rho_{\rm mono}$), but the final zero-temperature relaxation
of atomic coordinates and lattice vectors reduces the density somewhat,
such that the final Model I has a density $\sim$12\% smaller than
$\rho_{\rm mono}$.  A second amorphous structure, Model II, is generated
using Model I as the starting structure. The final mass density
increases $\sim$6\% in going from Model I to Model II.

Although the crystalline phases of ZrO$_2$ have been extensively investigated
both experimentally and theoretically (Ref.~\onlinecite{zro2} and
references therein), there has been very little work to characterize
the amorphous phase.  Experimental efforts have mainly involved
studies of $a$-ZrO$_2$ in the form of powders \cite{tanaka90,landron94} or  
thin films. \cite{winterer00}
As for a-ZrO$_2$ thin films, the resulting structure may be expected
to depend strongly on deposition and processing conditions and
on stoichiometric variations and impurities.
Moreover, while ZrO$_2$ films grown at high temperature are
typically polycrystalline, there are hints that ZrO$_2$ films grown at low
temperature may be a mixture of amorphous and polycrystalline
phases. One might also expect that the amorphous structure in thin films 
could be quite different from the ``bulk'' amorphous state that our
model attempts to simulate.
Thus, a direct comparison of our results with experiments may
not presently be feasible. In particular, Winterer \cite{winterer00}
used a thin-film sample (presumably a-ZrO$_2$) with a density
of 4.2\,g/cm$^3$, approximately 71\% of the density of
monoclinic ZrO$_2$ ($\rho_{\rm mono}$=5.89 g/cm$^3$). The density
of his sample corresponds to that of the 11-cell in our simulation.
According to our calculations, the volume at this density is too large to
sustain a void-free atomistic structure. On the
contrary, the densities of two amorphous structures we simulated
--- Model I and II --- are about 88\% and 94\% of $\rho_{\rm mono}$.
Such a large difference in density would naturally result in important
differences in the local structure (e.g., coordination numbers) and
the overall physical properties.
In any case, in view of the potential importance of ZrO$_2$ and
HfO$_2$ in CMOS gate dielectrics, it can be hoped that more
experimental work in this area may soon emerge.

\section{Dielectric Properties of \lowercase{$a$}-Z\lowercase{r}O$_2$}
\label{sec:diel}

\subsection{Introduction}
\label{sec:diel_intro}

The dielectric properties of amorphous ZrO$_2$ are calculated using the
linear-response features of the ABINIT package.  Since the structural
models generated in Sec.~\ref{sec:md_models} are
quite large and have essentially no symmetry,
the computation of their dielectric properties is quite
time-consuming.  Therefore, we could afford to carry out a full study of
the dielectric properties of only one of these models, and we somewhat
arbitrarily chose to focus on Model I.

Before computing the dielectric properties with ABINIT, the ground-state
structure of Model I obtained in VASP is first relaxed again using ABINIT
for consistency.  The lattice constant increases slightly from
10.717\,{\rm \AA} to 10.917\,{\rm \AA}, corresponding to a 5.6\% increase
in volume with respect to the ground-state volume of 1230.76\,${\rm \AA}^3$
obtained in VASP, while the atomic coordinates change very little
(less than 1\%). We use the newly relaxed structural
coordinates in our subsequent linear-response calculations. 

\subsection{Results}
\label{sec:diel_results}

The purely electronic dielectric tensor is computed to be
\def\phm{\phantom{-}}
\[\epsilon_{\infty} =  
 \pmatrix{ \phm4.76 &    -0.03 & \phm0.03 \cr
              -0.03 & \phm4.62 & \phm0.00 \cr
           \phm0.03 & \phm0.00 & \phm4.54 \cr}
          . \]
Clearly this tensor is approximately isotropic and diagonal,
as expected for any large supercell containing an amorphous material.
We obtain an orientationally-averaged dielectric constant of
$\epsilon_\infty$=4.6.
This is only slightly smaller than the values obtained previously
for the various crystalline phases of ZrO$_2$
($\sim\,$5.3--5.7),\cite{zro2,ortho,detraux98,rignanese01}
confirming our previous conclusion that $\epsilon_\infty$ is
fairly insensitive to the structural phase.

\begin{figure}
\begin{center}
   \epsfig{file=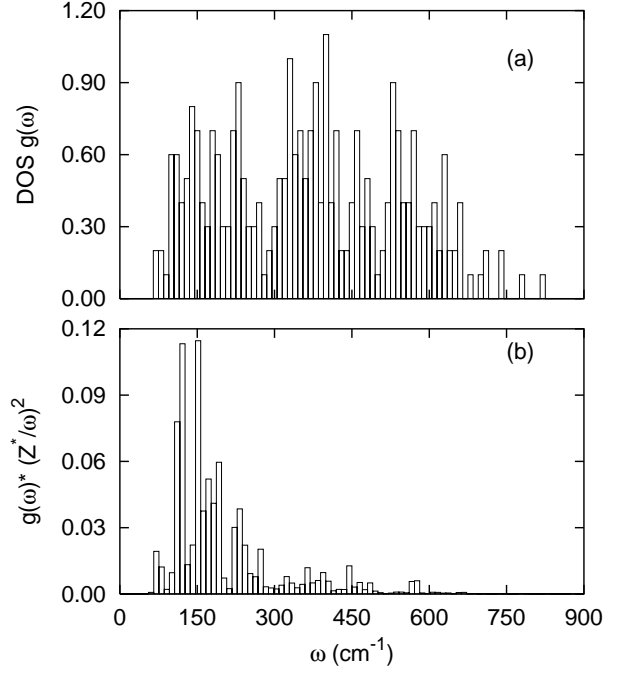,width=3.1in}
\end{center}
\caption{(a) Histogram of phonon density of states (DOS) vs.~frequency.
(Height of each bar is the number of phonon modes in the
bin divided by the bin width of 10\,cm$^{-1}$.)
(b) Histogram of DOS weighted by $\widetilde{Z}_{\lambda}^{*\,2} /
{\omega}_{\lambda}^{2}$.}
\label{fig:diel_dos}
\end{figure}

In order to obtain the lattice contribution to the dielectric tensor,
we first need to compute the phonon mode frequencies and effective
charges.  Beginning with the mode frequencies, we present in
Fig.~\ref{fig:diel_dos}(a) a histogram plot of the phonon DOS.
(To be more precise, this is the DOS of modes at the Brillouin zone
center of the supercell; this accounts for the absence of very
low-frequency modes in the plot.)
As can be seen in Fig.~\ref{fig:diel_dos}(a), the phonon modes extend up
to about 850 cm$^{-1}$, and aggregate roughly into three groups with
frequencies in the range 70 -- 300, 300 -- 510, and 510 -- 850 cm$^{-1}$.
The overall DOS spectrum does not show the kind of discrete features
expected for crystalline solids.

Next, we use the computed mode effective charges
${\widetilde{Z}}^{*}_\lambda$ and frequencies ${\omega}_{\lambda}$
of zone-center modes $\lambda$ to construct the ``infrared activity''
shown in Fig.~\ref{fig:diel_dos}(b) and defined as the phonon DOS weighted by
$\widetilde{Z}_{\lambda}^{*\,2} / {\omega}_{\lambda}^{2}$. (That
is, the frequency integral of this function gives the lattice
contribution to the isotropically averaged dielectric constant.  Here
the scalar mode effective charge ${\widetilde{Z}}^{*}_{\lambda \alpha}$,
defined via ${\widetilde{Z}}^{*\,2}_{\lambda} =
\sum_\alpha {\widetilde{Z}}^{*\,2}_{\lambda \alpha}$,
is essentially the projection
of the atomic $Z^*$ tensors onto the dynamical matrix eigenvectors as
described in Eqs.~(6-9) of Ref.~\onlinecite{zro2}.)
We find that the largest contribution to the lattice dielectric
response comes from modes of low frequency, $\sim$100-250$\,$cm$^{-1}$.

From these ingredients, we calculate the total lattice contribution
to the dielectric tensor, which is
\[\epsilon_{\rm latt} =
 \pmatrix{ \;20.3 & -1.2 & \;\;\;0.5 \cr
                -1.2 & \;17.1 & -0.4 \cr
          \;\;\;0.5  & -0.4 & \;15.5 \cr }
          . \]
Combining with the purely electronic dielectric tensor, the total
dielectric tensor is then
\[\epsilon_{0} =
 \pmatrix{ \;25.1 & -1.2 & \;\;\;0.5 \cr
                -1.2 & \;21.7 & -0.4 \cr
          \;\;\;0.5  & -0.4 & \;20.0 \cr }
          . \]
Once again, we find that these tensors are roughly diagonal and
isotropic, which tends to confirm that our structural model is
indeed amorphous.  Taking orientational averages, we can
summarize our results by observing that the purely electronic
dielectric constant $\epsilon_\infty$$\simeq$4.6 is augmented by a lattice
contribution $\epsilon_{\rm latt}$$\simeq$17.6 for a total dielectric
constant $\epsilon_0$$\simeq$22.  For the crystalline phases, we found
that $\epsilon_0$ and $\epsilon_{\rm latt}$ (in contrast to
$\epsilon_\infty$) are strong functions of crystal structure, with
$\epsilon_{\rm latt}$ ranging from about 14 to 33, or
$\epsilon_0$ ranging from about 19 to 39.\cite{zro2,hfo2,ortho,zrhf}
Here, we find that the dielectric constant of $a$-ZrO$_2$
($\epsilon_0$=22) is similar to that calculated previously for the
monoclinic ($\epsilon_0$=20) and two orthorhombic phases
($\epsilon_0$=20 and 19).\cite{zro2,ortho}

\subsection{Decomposition by atom type}
\label{sec:diel_decomp}

Clearly it is desirable to understand more fully the various
contributions to the lattice dielectric response of the amorphous
form of ZrO$_2$.  To this end, we now decompose various lattice
properties by ``atom type'' (that is, by chemical species and
coordination number) in the hope that such an analysis may provide
further insight into our numerical results.

\begin{figure}
\begin{center}
   \epsfig{file=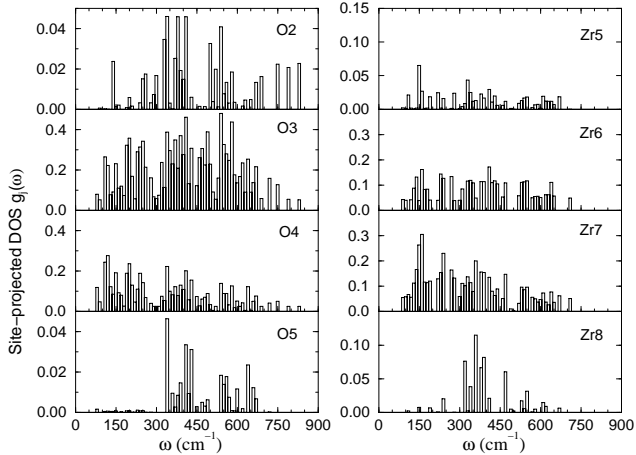,width=3.3in}
\end{center}
\caption{Site-projected phonon density of states $g(\omega)$ vs.~phonon
frequency for different atoms types (`O2' indicates 2-fold
coordinated oxygen, etc.) in amorphous Model I.}
\label{fig:ldos}
\end{figure}

We begin by decomposing the total phonon DOS $g(\omega)$ in
Fig.~\ref{fig:diel_dos}(a) into a local DOS on each type of atom
corresponding to the rows of Table \ref{table:md_bondlength}.
The total DOS
$g(\omega) = \sum_{\lambda} \delta(\omega_{\lambda} - \omega)$
can be decomposed as $g(\omega)=\sum_\tau g_\tau(\omega)$ where
the local DOS for atoms of type $\tau$ is
\begin{equation}
g_\tau(\omega) = \sum_{j \in \tau} \sum_{\lambda\alpha}
|e_{j \alpha}^{\lambda}|^2 \delta(\omega_{\lambda} - \omega).
\end{equation}
Here $\omega_{\lambda}$ is the eigenfrequency of phonon mode $\lambda$,
$\tau$ runs over atom types (O2--O5, Zr5--Zr8),
and $e_{j \alpha}^{\lambda}$ is the component of the eigenvector of
phonon mode $\lambda$ for atom $j$ along Cartesian direction
$\alpha$. The results are plotted in Fig.~\ref{fig:ldos}.

We were hopeful that such a decomposition might help us understand
whether the modes associated
with certain atom types are systematically much softer or harder.
Instead, no clear trends emerge from Fig.~\ref{fig:ldos}.
(The different appearance of the spectra for O5 and Zr8 most
probably results from the lack of statistics for these atoms,
which occur in the supercell only once and twice, respectively.)
A slight reweighting of the spectra towards softer mode frequencies
for increasing atomic coordination numbers can be observed, but
this is not a very strong effect.

\begin{figure}
\begin{center}
   \epsfig{file=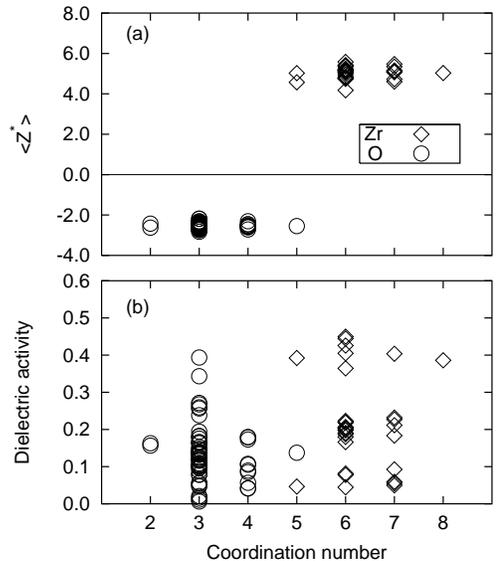,width=2.6in}
\end{center}
\caption{(a) Scatterplot of isotropically-averaged atomic $Z^*$
values (vertical axis) vs.~atom type and coordination number
(horizontal axis) for amorphous Model I.  Circles and diamonds
denote O and Zr atoms, respectively.  (b) Same but with `dielectric
activity' (see Eq.(\protect\ref{eq:activ})) plotted vertically.}
\label{fig:cn_diel_zstar}
\end{figure}

Next we test how the atomic Born effective charges correlate with
atom type.  Recall that the Born effective charge tensor
{\bf Z}$_{i,\alpha \beta}^*$ quantifies the macroscopic polarization
along Cartesian direction $\alpha$ induced by a displacement of
sublattice $i$ along $\beta$. Due to the amorphous nature of our model,
the calculated $Z^*$ tensors have no symmetry, although we find that
for most atoms the diagonal elements are dominant.  To reduce the
large quantity of $Z^*$ data to manageable proportions, we present in
Fig.~\ref{fig:cn_diel_zstar}(a) a scatterplot of just the isotropic
averages (i.e., one third of the trace) of the atomic $Z^*$ tensors
sorted by atom type for all 96 atoms.

We find that the $Z^*$ values are fairly tightly clustered about their
average values of 5.0 and $-$2.5 for Zr and O, respectively.
These are just slightly smaller than the typical values of 5.4 (Zr)
and $-$2.7 (O) computed for the crystalline phases,\cite{zro2,hfo2}
but still clearly larger than the nominal chemical valences of 4 and
$-$2, indicating that significant partial covalent character survives
in the amorphous phase.

To investigate the roles of the various atom types further,
we have carried out an analysis
in which we also make an atomic decomposition of the lattice dielectric
response.\cite{zro2} To do this, we first decompose the lattice dielectric
into contributions from pairs of atoms $i$ and $j$,
\begin{equation}
\epsilon_{\alpha\beta}^{\rm latt} = \sum_{ij}
\tilde{\epsilon}^{\,ij}_{\alpha \beta},
\end{equation}
where $\alpha$ and $\beta$ label the Cartesian directions. Here
\begin{equation}
\tilde{\epsilon}^{\,ij}_{\alpha \beta} =
\frac{4 \pi e^{2}}{V} \sum_{\lambda}
\frac{1}{\kappa_{\lambda}}R^{\lambda}_{\alpha i} R^{\lambda}_{\beta j},
\end{equation}
where $\kappa_{\lambda}$ and $e^{\lambda}_{j \beta}$ are the eigenvalue
and eigenvector of the force constant matrix $\Phi_{ij}^{\alpha \beta}$
for the phonon mode $\lambda$, $V$ is the volume of unit cell, and
$R^{\lambda}_{\alpha j} = \sum_{\beta} Z^{*}_{j, \alpha \beta}\,
e^{\lambda}_{j \beta}$.  We then heuristically define the contribution
coming from atom $i$ to be
\begin{equation}
\bar{\epsilon}^{\,(i)}_{\alpha \beta}=\sum_{j} \frac{1}{2}
\left( \tilde{\epsilon}^{\,ij}_{\alpha \beta} +
\tilde{\epsilon}^{\,ji}_{\alpha \beta} \right) \;.
\end{equation}
Finally, we will refer to the trace
\begin{equation}
\bar{\epsilon}_i=\sum_\alpha \bar{\epsilon}^{\,(i)}_{\alpha\alpha}
\label{eq:activ}
\end{equation}
as the ``dielectric activity" of atom $i$.

This quantity is plotted versus coordination number in the scatterplot of
Fig.~\ref{fig:cn_diel_zstar}(b).  The results indicate that a large
number of atoms contribute, but that a notable subpopulation of
mostly 6-fold Zr atoms are particularly strong contributors.

\subsection{Discussion}
\label{sec:discuss}

The dielectric constant ($\epsilon_0 = 22$) of our
amorphous model is much larger than that of SiO$_2$ ($\epsilon_0 =
3.5$) and quite comparable to the average dielectric constant of the
monoclinic phase.  Thus, we find that from the point of view of this
one criterion alone, $a$-ZrO$_2$ is indeed a promising high-$K$
dielectric for next-generation gate dielectrics.  Of course,
many other issues need to be addressed, not the least of which is
the stability (e.g., as measured by the recrystallization temperature)
of the amorphous phase.

Direct comparison with experiment is difficult, both because few
experimental measurements on amorphous ZrO$_2$ are available, and
because the sample preparation procedures may vary and may result
in rather different amorphous samples.\cite{devine01,liu04}  We
compare here with the
recent work of Koltunski and Devine, \cite{devine01} focusing
mainly on the sample deposited at room temperature using no rf bias
(for which spectra are shown in Figs.~1(c) and 2(c) of
Ref.~\onlinecite{devine01}) since the authors judged this sample to
be their ``most amorphous'' one.  (Upon annealing, the samples
tended to recrystallize partially to monoclinic and tetragonal
phases.) These authors reported a dielectric constant $\epsilon_0$
(at 100\,kHz) of $\sim$\,15 -- 18 as measured electrically on the
as-deposited ZrO$_2$ thin films.  The optical refractive index
measured at 632.8\,nm was in the range $n$ = 1.83 -- 1.85.  For this  
most amorphous sample, the absorbance spectrum, obtained using a
Nicolet Fourier-transform infrared spectrometer, shows some TO
modes beyond 1000\,cm$^{-1}$.  However, since our calculation does
not show any phonon modes in this range, we suspect these may be
defect-related features.  The sample shows broad TO and LO
features centered at frequencies identified as 410\,cm$^{-1}$ and
693\,cm$^{-1}$, respectively.\cite{devine01} (The authors also
pointed out that use of the Lyddane-Sachs-Teller relation would yield
an estimate of $\epsilon_{0} \simeq 10$ for the amorphous phase,
but this method substantially underestimates $\epsilon_{0}$
because the IR measurements did not extend to very
low frequency.)  While our value of 22 certainly exceeds the
experimentally measured one of 15--18, it is roughly the same order
of magnitude, and better agreement might not be expected in view
of the fact that the nanostructure and stoichiometry of the
experimental film are imperfectly characterized.

\section{Summary}
\label{sec:summary}

We have generated two realistic models of amorphous ZrO$_2$ via a
``melt-and-quench'' scheme using {\em ab-initio} MD simulations.
Candidate supercells of different volume were constructed in order
to study the effects of volume on the simulated structures. An
amorphous model (Model I) was obtained from one candidate structure
(i.e., the 7-cell), avoiding recrystallization or void formation
during the ``melt-and-quench'' sequence. The equilibrium density of this
model is $\sim$12\% less than that of the monoclinic phase.  The
second amorphous structure (Model II) was generated from Model I
with further processing.  We expect both models to be reasonable  
representatives of $a$-ZrO$_2$.  The structural properties and the
bond-length distributions of Model I and II were analyzed, and both
models were found to be composed mainly of 6- and 7-fold
coordinated Zr atoms and 3- and 4-fold coordinated O atoms.

The Born effective charges, phonon mode vectors and frequencies,
and electronic and lattice dielectric tensors were then calculated
for Model I using linear-response methods.  The phonon DOS was
found to be relatively featureless, and the infrared activity
spectrum (i.e., the DOS weighted by $\widetilde{Z}_{\lambda}^{*\,2}
/ {\omega}_{\lambda}^{2}$) showed a broad peak in the range of
100--250$\,$cm$^{-1}$ in
the phonon spectrum.  The calculations show that the Born
effective charges of the Zr and O atoms are fairly narrowly distributed
about 5.0 and $-$2.5, respectively, rather similar to what was
found previously for the crystalline phases.
Both the electronic and lattice contributions
to the dielectric tensor were found to be fairly isotropic,
as expected for an amorphous structure.  The calculated electronic
contribution was about 4.6, slightly less than for the
crystalline phases.  On the other hand, the lattice contribution to 
$\epsilon_{0}$ was calculated to be 17.6, rather similar to the
average value of monoclinic ZrO$_2$, though still significantly
smaller than for the cubic and tetragonal phases.

It is difficult to know just how similar our ``theoretical {\it 
a}-ZrO$_2$'' is to ``experimental {\it a}-ZrO$_2$''.  There are
difficulties on both sides of the comparison.
On the theoretical side, several approximations have been
introduced.  First, the modest size of our model supercell
(2$\times$2$\times$2) is a serious approximation, even though it is
essentially at the limit of current computational
capabilities.  However, the nearly isotropic form of the lattice
($\epsilon_{\rm latt}$) and electronic ($\epsilon_{\infty}$)
dielectric tensors tends to confirm that we have lost the memory of
the initial crystalline structure and obtained an amorphous one, in
spite of the supercell size limitation.  Second, recall that the
4$s$ and 4$p$ electrons were included in the valence in the Zr
pseudopotential for the MD simulations, but not for the
linear-response calculations.  This additional approximation could
affect the accuracy of the computed dielectric properties of our
amorphous structure.  However, we believe this approximation is not
very significant since the structural calculations on the
monoclinic phase of ZrO$_2$, which is rather ``disordered'' itself,
give nearly identical results using the two Zr pseudopotentials.
Experimentally, it is possible that the amorphous state manifests
itself differently in bulk and thin-film forms.  Indeed, because
ZrO$_2$ is essentially a poor glass-former, it can only be prepared
in amorphous form by low-temperature deposition or in other
non-equilibrium conditions, and it is therefore to be
expected that the properties of such samples may vary strongly with
preparation conditions.  Motivated by applications, much work is
now under way to explore whether chemical substitution (e.g., by
addition of Si, Al, or N) may help stabilize the amorphous phase
and raise its recrystallization temperature, and ultimately it will
be important to extend the theory to such substitutions in the
future.  Nevertheless, we believe that the present computational
investigation can serve as an important
first step in answering some of the many open questions about the
structural and chemical properties of this class of materials.

\section*{Acknowledgments}

This work was supported by  NSF Grant DMR-0233925. X.Z. thanks Jeff
Neaton for help with the VASP software package.

\end{document}